\documentclass[%
 reprint,
 amsmath,amssymb,
 aps,
]{revtex4-2}

\usepackage{graphicx}
\usepackage{dcolumn}
\usepackage{bm}
\usepackage{caption}
\usepackage{subcaption}

\usepackage{xcolor}

\begin{document}

\preprint{APS/123-QED}

\title{Generation of gravitating solutions with Baryonic charge from
Einstein-Scalar-Maxwell seeds}%

\author{Fabrizio Canfora$^{1,2}$}
\email{fabrizio.canfora@uss.cl}
\author{Anibal Neira$^{3}$}
\email{aneira2017@udec.cl}
\author{Seung Hun Oh$^{4}$}
\email{sho@tukorea.ac.kr}
\affiliation{$^{1}$Centro de Estudios Cientificos (CECs), Avenida Arturo Prat 514,
Valdivia, Chile.}
\affiliation{$^{2}$Facultad de Ingenieria, Arquitectura y Dise$\widetilde{n}$o, Universidad
San Sebastian, sede Valdivia, General Lagos 1163, Valdivia 5110693, Chile.}
\affiliation{$^{3}$Universidad de Concepci\'on (UDEC), Concepci\'on, Chile}
\affiliation{$^{4}$Department of Liberal Arts, Tech University of Korea,
Siheung-Si 15073, Korea}


\begin{abstract}
We establish, for the first time, an exact correspondence between Einstein–scalar–Maxwell theory and gauged Skyrme–Maxwell–Einstein models in $(3+1)$ dimensions. By constructing the simplest consistent ansatz within the gauged Skyrme–Maxwell framework, we reveal a remarkable equivalence in a sector that admits nonvanishing, highly magnetized baryonic charge.
This correspondence has a particularly appealing consequence: it transfers the full power of solution-generating techniques developed for electrovacuum systems—many of which naturally accommodate scalar fields—to the considerably more intricate setting of gauged Skyrme–Maxwell theory minimally coupled to General Relativity. As a result, it opens the door to a systematic and much broader exploration of exact solutions in Skyrme–Maxwell–Einstein theory and of their potential applications in cosmology and astrophysics. Notably, the resulting configurations carry nonzero baryonic charge whenever the derivative of the hadronic profile along the magnetic field lines does not vanish.
As an illustrative example, we apply this new dictionary to a rotating Kerr–Newman–like spacetime dressed with a scalar field. In the corresponding Skyrme–Maxwell–Einstein solution, the quantization of the baryonic charge enforces a quantization of the Kerr rotation parameter. We derive an upper bound on the baryonic charge in terms of the integration constants of the solution and show that, in the regime of small baryonic charge, the rotation parameter depends linearly on the baryonic charge.

\end{abstract}

\maketitle


\section{Introduction}

One of the most important—and at the same time most challenging—problems in General Relativity (hereafter GR), cosmology, and astrophysics is the study of baryonic charge dynamics in regimes characterized by strong gravitational fields, rapid rotation, and intense magnetic fields (see, e.g., \citep{Broderick,Sathyaprakash,Giataganas,Watanabe,Berryman}).

In this regime, numerical simulations become extremely demanding, while available analytical techniques are often ineffective, as one must confront the intrinsic nonlinearity of GR coupled to another highly nontrivial theory—the low-energy limit of QCD (hereafter LEQCD). The latter is notoriously difficult to handle: neither lattice QCD nor perturbative methods perform reliably in this domain (see \cite{sign1, sign2, sign3m, HIC, QGBook, Manton, Shuryak, Shifman, Kogut} and references therein), since LEQCD is dominated by genuinely non-perturbative effects \cite{Forcrand, Brambilla}.
Nevertheless, LEQCD admits an effective description in terms of $(3+1)$-dimensional gauged Skyrme–Maxwell theory (GSMT) in the presence of $SU(2)$ isospin symmetry (see \cite{Manton, Skyrme1, Skyrme2, Skyrme3, Skyrme4, BaMa} and references therein). GSMT plays a central role within chiral perturbation theory (CPT; see \cite{CPT1, CPT2, CPT3, 1m, 2m, 3m, 4m}), where the topological charge density is naturally interpreted as baryonic charge density.

Developing effective analytical tools in this framework would be highly desirable—not only to complement and guide numerical simulations, but also to uncover novel phenomena arising from highly magnetized baryonic matter in regions of strong gravitational fields.

Here we accomplish this goal by exhibiting a remarkably simple exact mapping between the Einstein–scalar–Maxwell theory and the gauged Skyrme–Maxwell theory minimally coupled to GR in $(3+1)$ dimensions. The mapping relies on the simplest ansatz in the GSMT that preserves a non-vanishing baryonic charge density. This construction has potentially far-reaching implications, ranging from cosmology to astrophysics (which we leave for future investigations).

The present manuscript concentrates on a particularly intriguing consequence of this mapping: \textit{the teleportation of solution-generating techniques developed in the scalar electrovacuum via the systematic transfer of their outcomes to the Skyrme–Maxwell–Einstein theory, thereby constructing exact solutions with non-vanishing baryonic charge}. This is the first framework in which baryonic degrees of freedom are combined in a genuinely non-trivial way with solution-generating methods of GR, which by themselves were previously unable to produce configurations carrying baryonic charge. The result is especially significant, as it enables a controlled analysis of baryonic charge distributions in regions of strong gravity, intense magnetic fields, and rapid rotation, among other possibilities. 

The present approach exploits the structure of the baryonic charge density in the presence of minimal electromagnetic coupling, which splits into the standard Skyrme and Callan–Witten terms. We adopt the simplest ansatz with the desired properties, for which the Skyrme contribution vanishes identically, leaving the baryonic charge density entirely supported by the Callan–Witten term. Consequently, this formalism describes intrinsically highly magnetized baryonic matter: when the magnetic field is switched off, the baryonic charge density vanishes as well.

This paper is organized as follows. Section II reviews the Einstein–gauged Skyrme–Maxwell theory, emphasizing the two distinct contributions to the baryonic charge density: the standard Skyrme term and the Callan–Witten term. In Section III, we introduce an exact mapping based on a specific ansatz for the $SU(2)$ chiral field, which reduces the highly nonlinear equations of the gauged Skyrme–Maxwell theory to the considerably simpler Einstein–Maxwell–scalar system. In Section IV, this dictionary is applied, showing how analytic rotating solutions with non-vanishing baryonic charge density can be generated. Finally, Section V presents our concluding remarks and discusses possible implications for future astrophysical and cosmological investigations.

\section{The gauged Skyrme-Maxwell theory}

The action of the gauged Skyrme–Maxwell–Einstein theory is given by (see
\cite{Skyrme1, Skyrme2, Skyrme3, Skyrme4, Manton, BaMa, 1m, 2m, 3m, 4m})
\begin{widetext}
\begin{align} S=\int d^{4}x \sqrt{-g} \, \bigg\{ \frac{1}{4} R -\frac{1}{4}F_{\mu\nu} F^{\mu\nu}
+\frac{K}{4}\mathrm{Tr}\left[ L^{\mu}L_{\mu}+\frac{\lambda} {8}G_{\mu\nu}G^{\mu\nu}\right] \bigg\} \, ,\label{ESMaction}
\end{align}
\end{widetext}
where $R$ denotes the Ricci scalar and $K=(f_{\pi})^{2}/4$ is the Skyrme coupling constant, experimentally fixed to $f_{\pi}\approx130\ \text{MeV}$. Throughout this work we adopt the conventions $c=4\pi G=\epsilon_{0}=1$. The pion mass $m_{\pi}$ is neglected, as we focus on configurations with large baryonic charge, for which $m_{\pi}$ is much smaller than the characteristic energy scales involved.

The field equations for the gravitational part read
\begin{align}
R_{\mu\nu} - \frac{1}{2}R g_{\mu\nu} = 2 T_{\mu\nu}\ , \label{EinsteinEqu1}%
\end{align}
where $R_{\mu\nu}$ is the Ricci tensor while
\begin{widetext}
\begin{equation}
T_{\mu\nu}=  -\frac{K}{2}\operatorname{Tr}\bigg[ L_{\mu}L_{\nu
}-\frac{1}{2}g_{\mu\nu}L_{\sigma}L^{\sigma}
 +\frac{\lambda}{4}\left(  g^{\alpha\beta}G_{\mu\alpha}G_{\nu\beta}-\frac
{1}{4}g_{\mu\nu}G_{\alpha\beta}G^{\alpha\beta}\right)  \bigg]
 + F_{\mu\alpha}F_{\nu}^{\;\alpha}-\frac{1}{4}F_{\alpha\beta}F^{\alpha\beta} g_{\mu\nu}\ . \label{tmunu(1)}
\end{equation}
\end{widetext}
In addition
\begin{align}
L_{\mu}  &  =U^{-1}D_{\mu}U=L_{\mu}^{j}t_{j},\ F_{\mu\nu}%
=\partial_{\mu}A_{\nu}-\partial_{\nu}A_{\mu}\,,\nonumber\\
G_{\mu\nu}  &  =\left[  L_{\mu},L_{\nu}\right]  \ ,\ t_{j}%
=i\sigma_{j}\,,
\end{align}
where $\sigma_{i}$ are the Pauli matrices and the $SU(2)$ valued $U$ field can
be parametrized as
\begin{align}
U  &  =1_{2\times2}\cos\Psi+\left(  \sin\Psi\right)  n^{j}t_{j}%
\,,\label{eq:defSigma}\\
\overrightarrow{n}  &  =\left(  \sin\Theta\cos\Phi,\sin\Theta\sin\Phi
,\cos\Theta\right), \nonumber
\end{align}
where $1_{2\times2}$\ is the identity $2\times2$ matrix while $\Psi\left(
x^{\mu}\right)  $, $\Theta\left(  x^{\mu}\right)  $ and $\Phi\left(  x^{\mu
}\right)  $ are the three scalar degrees of freedom of the $SU(2)$-valued
Chiral field. The gauge-covariant derivative is defined as
\begin{equation}
\ D_{\mu}U=\partial_{\mu}U+A_{\mu}\left[  t_{3},U\right]  \ . \label{sky2.75}%
\end{equation}
Now, the field equations for the GSMT yield
\begin{align}
D_{\mu}\left(  L^{\mu}+\frac{\lambda}{4}[L_{\nu},G^{\mu\nu}]\right)
=0\ , \quad\nabla_{\mu}F^{\mu\nu}=J^{\nu}\ , \label{eq:NLSM}%
\end{align}
where $\nabla_{\mu}$ is the Levi-Civita covariant derivative, and the current
$J^{\mu}$, with $\widehat{O}=U^{-1}t_{3}U-t_{3}$, is given by
\begin{equation}
J^{\mu}=\frac{K}{2}\text{Tr}\left[  \widehat{O}\left(  L^{\mu}%
+\frac{\lambda}{4}[L_{\nu},G^{\mu\nu}]\right)  \right]  \ .
\label{maxcurrent}%
\end{equation}
Even in the absence of gravitational coupling, the field equations of the gauged Skyrme–Maxwell theory are already highly non-trivial. One might therefore expect that coupling the system to GR would further complicate the analysis; as we will show, however, this is not the case.

The corresponding Baryonic charge~\cite{Skyrme1,Skyrme2,Skyrme3, Skyrme4, Bala1, Witten,
gaugesky1} is given by
\begin{equation}
Q_B=\frac{1}{24\pi^{2}}\int_{S}\rho_{B}\ ,\ \ \rho_{B}=\rho_{B1}+\rho_{B2}\ ,
\label{new4.1}%
\end{equation}
where $S$ is a spacelike three-dimensional hypersurface while
\begin{align}
\rho_{B1}  &  =\epsilon^{ijk}\text{Tr}\left\{  \left(  U^{-1}\partial
_{i}U\right)  \left(  U^{-1}\partial_{j}U\right)  \left(  U^{-1}\partial
_{k}U\right)  \right\}  \ ,\label{rhoSk}\\
\rho_{B2}  &  =-3\epsilon^{ijk}\text{Tr}\left\{  \partial_{i}\left[
A_{j}t_{3}\left(  U^{-1}\partial_{k}U+\left(  \partial_{k}U\right)
U^{-1}\right)  \right]  \right\}  \,, \label{rhoMax}%
\end{align}
are the two topological density contributions, $A_{j}$ denoting the spatial
components of the gauge potential.

When minimal coupling to the Maxwell field is taken into account—a requirement in many physically relevant situations—a remarkable possibility emerges: \textit{one can construct a topologically non-trivial ansatz for which the standard Skyrme contribution $\rho_{B1}$ vanishes identically, while the Baryonic charge is entirely supported by the Callan–Witten term, $\rho_{B2}\neq0$ \cite{gravtube}}. This observation makes it possible to formulate a consistent ansatz for the hadronic and electromagnetic degrees of freedom that significantly simplifies the analysis of gravitating configurations carrying Baryonic charge in the presence of magnetic fields and rotation.

\section{The Mapping}

To incorporate Baryonic charge density in the analysis of configurations involving strong gravitational and magnetic fields, as well as rotation, we consider the following ansatz for the $SU(2)$-valued Skyrme field $U$ and the gauge field $A$
\begin{align}
\Psi & =\Psi(x^{\mu}),\quad  \Theta=\pi ,\quad \Phi=0 \,,\quad U=\exp\left(
\Psi t_{3}\right) , \nonumber\\
 A_{\nu}&=A_{\nu}(x^{\mu})\label{LWP0}.
\end{align}
The spacetime line element will be specified later, once the class of solutions employed in our dictionary has been chosen.

For the ansatz introduced above, the matter field equations of the gauged Skyrme-Maxwell-Einstein theory simplify drastically and reduce to
\begin{align}
\nabla_{\mu}F^{\mu\nu} & =0,\quad  \nabla_{\mu}\nabla^{\mu}\Psi=0\ , 
\label{LWP2}%
\end{align}
namely, to the sourceless Maxwell and Klein-Gordon equations. 
The only restriction stems from the fact that the GSMT represents the low-energy effective description of QCD and therefore ceases to be valid once the energy density associated with the hadronic degrees of freedom becomes too large. In the present context, this requirement translates (in natural units) into the condition
\begin{equation}
\left\vert \nabla_{\nu}\Psi\right\vert \lesssim1 \text{GeV} \ . \label{LQCD}%
\end{equation}
The origin of the remarkable simplification appearing in Eq. (\ref{LWP2}) lies in the fact that, for the ansatz introduced in Eq. (\ref{LWP0}), the $U(1)$ current $J^{\mu}$ vanishes identically. This follows from the observation that
\begin{equation}
\widehat{O}=\left( \exp\left( -\Psi t_{3}\right) \right) t_{3}\exp\left(
\Psi t_{3}\right) -t_{3}=0.
\end{equation}
As a consequence, no additional contributions arise from the gauged Skyrme sector, leading directly to the reduced field equations in Eq. (\ref{LWP2}).

Moreover, all commutators appearing in the Skyrme field equations vanish identically, so that the dynamics of the profile reduces to a linear equation, fully decoupled from the Maxwell sector. At first sight, this might suggest that the system belongs to a trivial sector of the theory. This conclusion, however, is misleading.
Indeed, the Callan–Witten contribution to the Baryonic charge density remains non-vanishing and is given by
\begin{align} \rho_{B}= & -3\epsilon^{ijk}\text{Tr}\left\{ \partial_{i}\left[ A_{j} t_{3}\left( U^{-1}\partial_{k}U+\left( \partial_{k}U\right) U^{-1}\right) \right] \right\} ,\nonumber\\ 
\approx & \overrightarrow{B}\cdot\overrightarrow{\partial}\Psi\approx B^{i}\partial_{i}\Psi\ .\label{LWPcharge}
 \end{align}
 
Note that $\rho_{B}$ admits the interpretation of a Baryonic charge density only when the three indices $i$, $j$, and $k$ appearing in $\epsilon^{ijk}$ are all spacelike. Consequently, whenever the derivative of the Skyrme profile $\Psi$ along the magnetic field $\overrightarrow{B}$—namely $B^{i}\partial_{i}\Psi$—is non-vanishing, the Baryonic charge density is non-zero as well.
This observation has an important implication: all existing results obtained within the theory of Einstein-scalar-Maxwell can be directly imported and reinterpreted in the context of the GSMT minimally coupled to GR. Naturally, the physically relevant configurations within this correspondence are precisely those for which $\rho_{B}\neq0$.

It is useful to revisit the following line of reasoning. As has been discussed in the literature, sub-leading corrections to the Skyrme model arise naturally in the ’t Hooft expansion (see \cite{Subleading1, Subleading2, Subleading3, Subleading4, Subleading5, Subleading6}). These higher-order contributions, which are typically highly non-linear, could in principle undermine the simplicity of the present approach to embedding the GSMT within the Einstein–scalar–Maxwell framework.
Remarkably, this is not the case. One can show that, independently of how many sub-leading terms are included, the use of the ansatz in Eq. (\ref{LWP0}) always leads to the same field equations as those in Eq. (\ref{LWP2}). A proof of this statement is provided in the appendix of \cite{Subleading7}. This result highlights the robust and universal character of this sector of the theory.

The present mapping opens up a number of intriguing possibilities. For instance, it enables the study of the early cosmological evolution of the system described by Eq. (\ref{LWP2}). Within this ansatz, the hadronic profile $\Psi$ is completely decoupled from the Maxwell sector and evolves as a free scalar field. This feature allows for a particularly precise analysis of Baryonic charge density fluctuations through Eq. (\ref{LWPcharge}), making transparent the direct relationship between magnetic field fluctuations and Baryonic charge fluctuations. We will return to this interesting issue in a future publication. In the present manuscript, however, we focus on a different non-trivial outcome of the analysis.

\section{An application of the Mapping: Baryonic degrees of freedom from a Kerr-Newman-like black hole with scalar dress}

Here we consider a Kerr–Newman–like black hole endowed with a nontrivial scalar field configuration. Through our mapping, this configuration generates a novel solution of the gauged Skyrme–Maxwell–Einstein theory carrying a nonvanishing Baryonic charge. The magnetic field of the original seed solution is appropriately aligned with the gradient of the scalar dressing, thereby ensuring the nontriviality of the resulting Baryonic charge profile.

Although we focus on a specific example, the scope of the construction is far more general. Owing to the fact that the backreacting gauged Skyrme–Maxwell theory maps into the Einstein–scalar–Maxwell system, the space of admissible seed solutions capable of generating configurations with nonvanishing Baryonic charge is vastly enlarged. Indeed, the Einstein–scalar–Maxwell system is known to possess a high degree of integrability. This is exemplified by the Ernst formulation of the electrovacuum \cite{Er1}–\cite{Er4}, as well as by other symmetry-based techniques \cite{Er5}.

The key idea  is that, in the presence of suitable symmetries, the electrovacuum (or scalar–electrovacuum) field equations admit a dimensional reduction along the associated Killing vectors. This reduction reveals a set of otherwise hidden Lie point symmetries that can be exploited to systematically generate new solutions. The availability of our dictionary makes it possible to export these powerful methods directly to the gauged Skyrme–Maxwell sector, thereby providing access to a wide class of solutions with nonvanishing Baryonic charge that would be practically unattainable through a direct integration of the GSMT field equations. We will report soon on these constructions. 

Hence, we shall focus on axisymmetric spacetimes that are conveniently described by the Weyl-Lewis-Papapetrou metric, which admits two commuting Killing vector fields, namely $\partial_{t}$ and $\partial_{\varphi}$. In the context of Einstein–Maxwell theory, this class of spacetimes is particularly amenable to analytic treatment due to the availability of powerful solution-generating techniques developed in \cite{Er1}–\cite{Er4} and \cite{Er5}. These methods provide an efficient way to control the nonlinear structure of the Einstein–scalar–Maxwell equations, enabling the construction of physically relevant solutions that would otherwise be difficult to uncover, especially when rotational effects are taken into account.

Notably, these techniques have led to the discovery of analytic black hole solutions embedded in magnetic universes \cite{Er18,Er19}, as well as rotating multi–black–hole configurations. This line of research remains highly active, with several recent and exciting developments reported in \cite{Astorino,Barrientos25,Barrientos:2023tqb,Cisterna:2023uqf,Barrientos:2023dlf,Barrientos:2024pkt,Barrientos:2024uuq,Barrientos:2025rjn}.

Hence, we consider the following spacetime configuration 
\begin{align}
& ds^{2} =-f\left( dt-\omega d\varphi\right) ^{2}+\frac{1}{f}\left[
\rho^{2}d\varphi^{2}+e^{2\gamma} \left( d\rho^{2}+dz^{2}\right)
\right], \\
& A_{\mu}dx^{\mu} =A_{t}(\rho,z)dt+A_{\varphi}(\rho,z)d\varphi
\ ,\label{appliernst1}\\
&\Psi=\Psi(\rho,z), 
\end{align}
where the metric functions $f$ and $\gamma$, as well as the scalar profile $\Psi$ (basically the Skyrme field in our dictionary), depend exclusively on the coordinates $\rho$ and $z$, as dictated by the stationarity and axisymmetry of our seed. The seed, to be explicitly shown next, precisely fits under this family. We will see its explicit form and how it induces a nontrivial baryonic charge in the backreacting GSMT side.

Hence, the seed solution to be employed was introduced in \cite{Barrientos25}. The metric is taken to be that of a Kerr–Newman–like spacetime, supplemented by a conformal factor $H(r,\theta)$ multiplying the spacelike transverse section at constant $t$ and $\phi$. This conformal factor fully captures the backreaction of the scalar field, as can be understood from the Eris–Gürses theorem \cite{Eris:1976xj}
\begin{align}
ds^{2} &  =-\frac{\Delta}{\rho^{2}}\big(dt-a\sin^{2}\theta\,d\varphi
\big)^{2}+\rho^{2}H\bigg(\frac{dr^{2}}{\Delta}+d\theta^{2}\bigg)\nonumber\\
&  +\frac{\sin^{2}\theta}{\rho^{2}}\big(adt-(r^{2}+a^{2})d\varphi\big)^{2}\ ,
\end{align}
where $\Delta=r^{2}-2Mr+a^{2}+e^{2}$, and $\rho^{2}=r^{2}+a^{2}\cos^{2}\theta$. Here, $M$ and $e$ denote the mass and electric charge of the system, respectively, while $a$ is the angular momentum parameter. We emphasize that spherical-like coordinates $(r,\theta)$ are used instead of the cylindrical-like coordinates $(\rho,z)$ adopted previously.
The Maxwell field corresponds to that of a Kerr–Newman black hole,
\begin{equation}
A=-\frac{er}{\rho^{2}}dt+\frac{aer\sin^{2}\theta}{\rho^{2}}d\varphi\ .
\end{equation}
As is well known, the scalar backreaction decouples from the electromagnetic sector.

\begin{figure}[t]
\includegraphics[width=.45\textwidth]{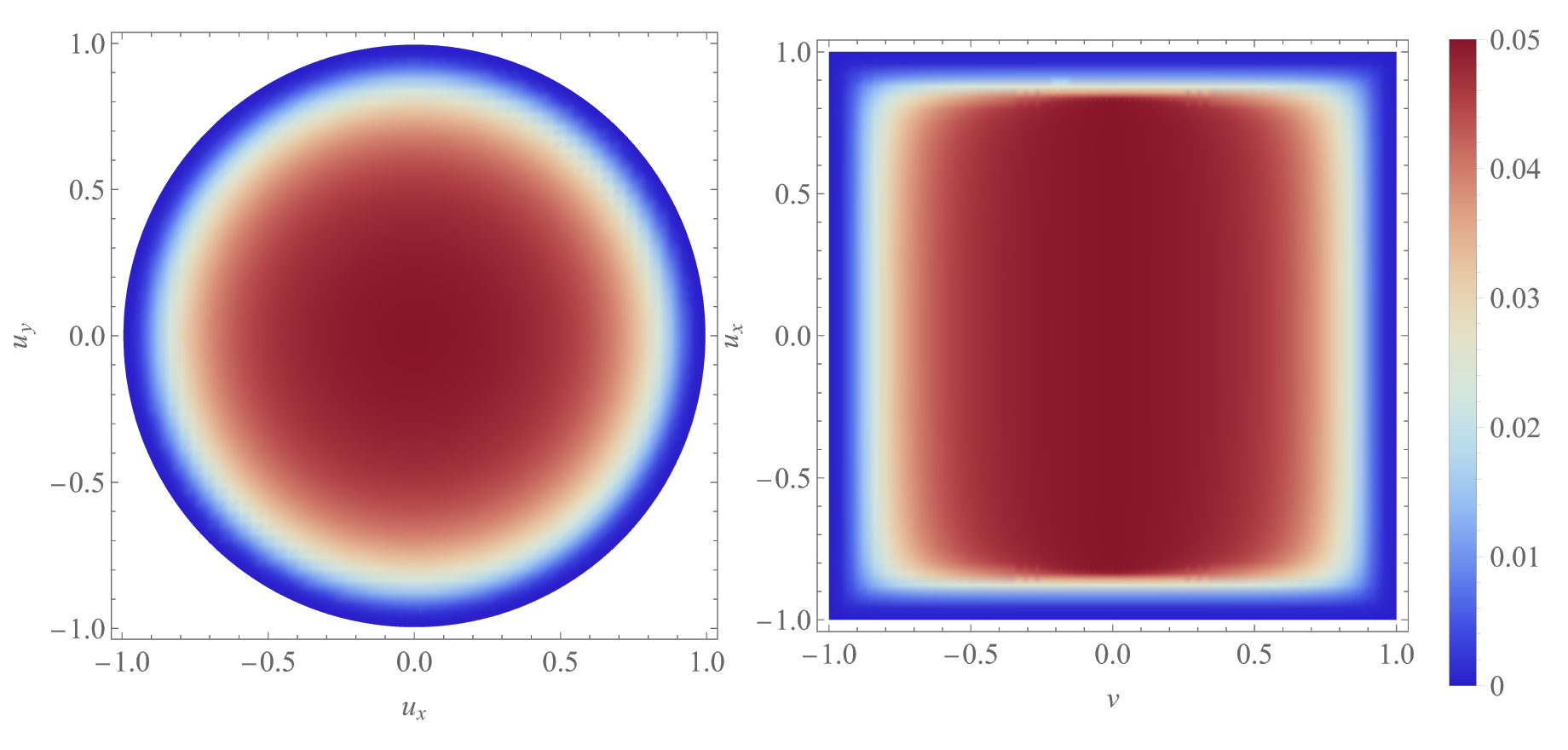}\caption{spatial distribution of the baryonic charge density $\rho_B$ (units of cm$^{-3}$) where  we have used stereographic coordinates to represent the concentration of the charges such that the center of the graphics represent the outer horizon and the borders represents spatial infinity. We have used $\Theta_0=180$, $M=4$, $e=1$ and $a=1$.} %
\label{fig:density}%
\end{figure}

At spatial infinity, the Maxwell field asymptotically approaches a dyonic configuration, consisting of the electric potential of a point charge together with the magnetic potential of a Dirac magnetic monopole. The Skyrme (scalar) profile $\Psi$ is characterized by two charge-like parameters, denoted by $\Sigma$ and $\Theta_0$, and takes the form of
\begin{equation}
\Psi=\frac{\Sigma}{2e}\ln\bigg(\frac{r-r_{+}}{r-r_{-}}\bigg)+\frac{\Theta_0}
{2e}\ln\bigg(\frac{1-\cos\theta}{1+\cos\theta}\bigg)\ ,
\end{equation}
where $r_{\pm}=M\pm\sqrt{M^{2}-a^{2}-e^{2}}$. The physical interpretation of the parameter $\Theta_0$ emerges only within the present framework, as we now explain. As mentioned, the gravitational backreaction induced by the scalar field $\Psi$ is fully encoded in the conformal factor $H$, which is given by
\begin{align}
H(r,\theta) &  =\bigg(1+\frac{(r_{+}-r_{-})^{2}}{4\Delta}\sin^{2}%
\theta\bigg)^{-K\Sigma^{2}/e^{2}}\nonumber\\
&  \times\bigg(\frac{(r_{+}-r_{-})^{2}}{4}+\frac{\Delta}{\sin^{2}\theta
}\bigg)^{-K\Theta_0^{2}/e^{2}}\nonumber\\
&  \times\bigg(\frac{2(r-M)-(r_{+}-r_{-})\cos\theta}{2(r-M)+(r_{+}-r_{-}%
)\cos\theta}\bigg)^{2K\Sigma\Theta_0/e^{2}}\ .
\end{align}
With these ingredients at hand the corresponding Baryonic charge density $\rho_{B}$ can be easily computed, yielding
\begin{align}
\rho_{B} &  =\frac{4a\Sigma}{r^2\Delta\rho^{4}}(r_+-r_-)(r^2+a^2) r \cos\theta\nonumber\\
&  +\frac{a\Theta_0}{r^2\rho^{4}}\big(r^{2}-a^{2}\cos^{2}\theta\big)\,\ .
\end{align}
The first term in $\rho_{B}$ does not contribute to the net baryonic charge, as its dependence on $\theta$ causes it to integrate to zero. By contrast, the second term yields a nontrivial contribution, entirely fixed by the parameters $M$, $a$, and $\Theta_0$, and clearly exhibits the factorized structure of the baryonic charge density.
A typical plot of  topological charge densities of this system is presented in
Fig. (\ref{fig:density}). The net baryonic charge is given by 
\begin{equation}
n=\pm\Theta_0\,\bigg[1-\frac{2}{\pi}\arctan\bigg(\frac{r_{+}}{|a|}%
\bigg)\bigg]\ ,\qquad n\in\mathbb{Z}\ ,
\end{equation}
where $\pm$ is chosen to correspond with the sign of $a$. Thus, the
charge-like parameter $\Theta_0$ gives an upper bound of $|n|$ so that the
system admits \textit{finitely} many baryonic charges only. 

\begin{figure}[t]
\includegraphics[width=.45\textwidth]{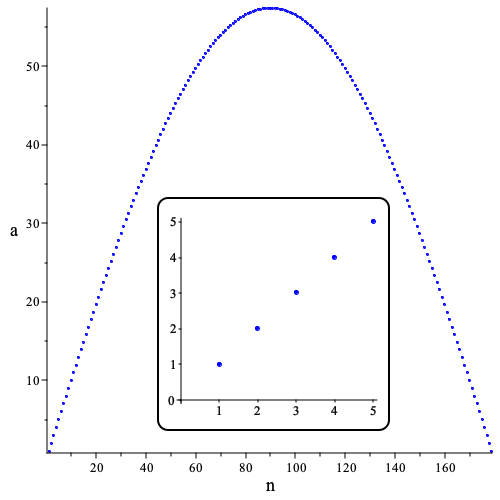}\caption{Discrete spectrum of
the specific angular momentum $a_{n}^{+}$ (units of $a_{\odot}$) as a function
of the baryonic charge $n$. Parameters are fixed at $M=M_{\odot}$,
$e=e_{\text{max}}$, and $\Theta_0=180$. The inset highlights the linear growth
of $a_{n}^{(+)}$ for small $n$ values.}%
\label{fig:t_charge}%
\end{figure}

Now, we present a quantization condition of the specific angular momentum
\begin{equation}
a_{n}^{(\pm)}=\frac{M}{2}\sin\bigg(\frac{\pi n}{\Theta_0}\bigg)\bigg[1\pm
\sqrt{1-\frac{e^{2}}{M^{2}}\sec^{2}\bigg(\frac{\pi n}{2\Theta_0}\bigg)}\bigg]\ .
\end{equation}
There is a direct consequence of this quantization which is a maximum value of the angular momentum, that is reached at $n_{max}=\frac{2\Theta_0}{\pi} \cos^{-1}\left( \frac{M}{\sqrt{2M^2-e^2}}\right)$ and the maximum value is then 
\begin{align}
    a_{max}^{+}=&\frac{M}{2} \left( 1+\sqrt{\frac{(M^2-e^2)^2}{M^4}}\right) \nonumber\\
    &\times \sin \left( 2 \cos^{-1}\left( \frac{M}{\sqrt{2M^2-e^2}}\right)\right)
\end{align}
Notice that $|a_{max}^+|<M$.
In our convention, the mass and the specific angular momentum of the sun are
$M_{\odot}\approx1.85\times10^{4}\,\text{m}$, and $a_{\odot}\approx
322\,\text{m}$, respectively. Moreover, a theoretical upper bound for the net
electric charge of a main-sequence star like the Sun is approximately
$e_{\max}/M_{\odot}\approx5.84\times10^{-19}$ \cite{maxcharge}. In this case,
$|a_{n}^{(-)}|$ is suppressed and the dominant contribution of the angular
momentum comes from $a_{n}^{+}$, which is presented in Fig.
(\ref{fig:t_charge}). $a_{n}^{(+)}$ grows linearly for sufficiently small $n$
\begin{equation}
a_{n}^{(+)}\approx\bigg(\frac{\pi M}{\Theta_0}\bigg)n\ .
\end{equation}

Note that, when $\Sigma=0$, $\rho_{B}$ is regular and well behaved even if
$\nabla_{\mu}\Psi$ is not. The condition in Eq. (\ref{LQCD}) applied to the
present rotating solution implies that the solution for the Hadronic
profile cannot be trusted when the distance from the outer horizon is less
than 1 Fermi (approximately). The baryonic charge of the system is obtained
integrating $\rho_{B}$ from the outer horizon up to spatial infinity.

\section{Final remarks}

We have constructed, for the first time, an exact dictionary between the Einstein–Maxwell–scalar theory and the gauged Skyrme–Maxwell theory minimally coupled to General Relativity in $(3+1)$ dimensions, a correspondence that remains unaffected by subleading corrections to the Skyrme model in the ’t Hooft large-$N_{c}$ expansion. We identified the simplest possible ansatz that reveals this remarkable relation within a sector admitting non-vanishing Baryonic charge.
This mapping has significant implications, as it allows a broad class of exact solutions of the Einstein–scalar–Maxwell system to be systematically promoted to exact solutions of the gauged Skyrme–Maxwell–Einstein theory. Potential applications range from cosmology to astrophysics, including, for instance, the study of correlations between Baryonic charge and magnetic-field fluctuations. As a concrete example, we constructed the first analytic rotating solution in gauged Skyrme–Maxwell–Einstein theory, in which the requirement of an integer Baryonic charge leads to a nontrivial quantization condition.

Overall, this framework opens a wide range of new possibilities for the exact analysis of Baryonic degrees of freedom in regimes characterized by strong gravity, intense magnetic fields, and rotation, among other features that will be explored in future work.

\begin{acknowledgments}
The authors would like to warmly thank Adolfo Cisterna for illuminating discussions and suggestions. 
S. H. O. is particularly grateful to Dr. Miok Park for many insightful discussions. 
F. C. acknowledges support from Fondecyt grant No. 1240048 and Grant ANID EXPLORACION 13250014. 
A. N. is supported by ANID-Subdirección de Capital Humano/Doctorado Nacional/2025-21253071.
S. H. O. is supported by the National Research Foundation of Korea (NRF) funded by the Ministry of Science and Technology (Grant RS-2025-16071544). 
\end{acknowledgments}



\nocite{*}

\bibliography{apssamp}

\end{document}